\documentstyle[12pt,epsfig]{article}

\columnwidth\textwidth

\begin{document}
\pagenumbering{arabic}
\begin{titlepage}
\begin{flushright}
{BUTP-97/06}\\
{hep-lat/9702019}
\end{flushright}
\vspace{2cm}
\begin{center}
{\large\bf Pion loops in quenched Quantum Chromodynamics}\\
\vfill
{\bf Gilberto Colangelo$^{*}$ and Elisabetta Pallante$^{\#}$}
\\[0.5cm]
$^{*}$ INFN - Laboratori Nazionali di Frascati, Gruppo Teorico,\\
P.O. Box 13, I-00044 Frascati (Rome), Italy\\
$^{\#}$University of Bern, Sidlerstrasse 5,
CH-3012 Bern,\\ Switzerland\\[0.5cm]

\end{center}
\vfill
\begin{abstract}
We calculate the divergences of the generating functional of quenched
Chiral Perturbation Theory to one loop for a generic number of flavours.
The flavour number dependence of our result enlightens the mechanism of 
quark loop cancellation in the quenched effective theory for any Green 
function or $S$ matrix element.
We also apply our results to $\pi \pi$ scattering and evaluate the 
coefficient of the chiral log in the $S$--wave scattering lengths for 
the quenched case.
\end{abstract}
\vspace{1cm}
\hspace{1.2cm}PACS: 12.39.Fe, 12.38.Gc.
\vspace*{1cm}
\setcounter{equation}{0}
\setcounter{figure}{0}

\vfill
 February 1997
\end{titlepage}

\everymath={\displaystyle}
\vspace*{-2mm}
Lattice simulations are at present the only available method to study 
in a quantitative way the nonperturbative regime of Quantum
Chromodynamics. Although theoretically well founded, the method is
limited by a number of technical problems, and by the computing power
which is not yet enough to fully cope with QCD, despite continuous
progress. 
One of the main technical difficulties is the implementation of an
efficient algorithm to evaluate the path integral over fermions, which
is at present very time consuming. The usual way to go around this 
problem has been the use of the ``quenched'' approximation, i.e. setting 
to one the determinant resulting from the integration over fermions. 
In the language of Feynman diagrams, this approximation corresponds to 
neglecting quark loops.

Quenching is so widely used that it has become important to understand 
{\em a priori} and in a quantitative way what is the effect introduced 
by this approximation \cite{first}. A very promising approach to this 
problem is, in our opinion, the effective Lagrangian method
proposed by Bernard and Golterman \cite{qCHPT}. This is based
on the observation due to Morel (see Ref. \cite{first}) that the
quenched approximation to QCD can be described by a Lagrangian where
one adds to the usual quark fields a corresponding number of ghost
(complex) scalar fields, with the same kinetic term, and the same coupling 
to gluons. The ghost fields exactly cancel the fermionic determinant. 
When the quark masses are set to zero the quenched QCD Lagrangian has a 
larger symmetry than the usual $U(N)_L\times U(N)_R$, and
must have a special fermionic character, since it transforms bosons
into fermions. This extended symmetry is described by the graded group
$U(N|N)_L\times U(N|N)_R$. Following exactly the same steps that
lead from the globally symmetric QCD Lagrangian in the chiral limit to 
Chiral Perturbation Theory (CHPT) \cite{gl84,gl85}, Bernard and
Golterman have proposed an effective Lagrangian that should describe
the low energy regime of quenched QCD: quenched CHPT (often qCHPT in the
following). The basic new ingredient is the presence of 
ghost fermion and boson fields that enter only inside loops. These
are Goldstone particles of the extended symmetry, and can be viewed as
bound states formed by a quark and a ghost--antiquark, 
or viceversa, and ghost--quark ghost--antiquark. 
Also for these new fields the structure of the interaction is dictated
by the graded symmetry, which however leaves the coupling constants
unconstrained. The main advantage of this method is that one has in
principle a systematic way to study the modifications introduced by 
quenching: much as in CHPT, one can e.g. evaluate unambiguously the
chiral logs present in any observable quantity. 

\begin{figure}[ht]
\epsfxsize 11 cm
\epsfysize 3 cm
\begin{picture}(50,15) \end{picture}
\epsffile{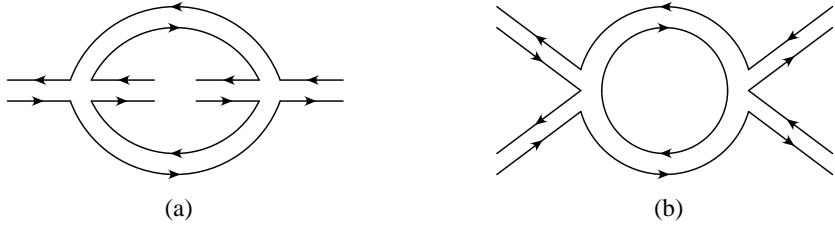}
\protect\caption{Two examples of pion loop graphs contributing to $\pi \pi$
scattering in the quark--flow diagram picture (all lines are quark lines). 
Diagram (a) does not contain quark loops, whereas diagram (b) does.}
\protect\label{fig1}
\end{figure}

Quenched CHPT has been successful in showing that the chiral logs
in the two point function of the axial current to one loop disappear
(i.e. both in $F_\pi$ and $M_\pi$, see Ref. \cite{qCHPT}), and that the
anomalous $\eta^{\prime}$ field is ill--defined, developing a double
pole in the propagator, as one would expect. The systematics of the 
cancellation of chiral logs for two point functions has been further 
investigated in Ref. \cite{pqCHPT}.
However, a general and strong argument in favour of the validity
of this effective Lagrangian method is still missing.
Such an argument could be produced if one were able to prove that the
cancellation mechanism between pion and ghost loops is such that the
pion loops that do not contain quark loops survive it (e.g. in $\pi
\pi$ scattering the graph in Fig. \protect\ref{fig1}a should remain
whereas that in Fig. \protect\ref{fig1}b not). 
This should be done in general, for any Green function or S matrix
element. 

Although at first sight this may seem difficult, there is in fact a
straightforward way to do it. The idea is to develop a path integral
formulation of quenched CHPT much on the same line as that of ordinary
CHPT. There, Gasser and Leutwyler \cite{gl85} have shown that, by
using the background field method and heat kernel techniques, it is
possible to derive in closed form the divergent part of the one loop
generating functional. This is a chiral invariant Lagrangian at
order $p^4$. The calculation was done in a general $SU(N)_R
\times SU(N)_L$ theory, so that the $N$ dependence of the divergent
part of the one loop generating functional is explicit. In the
result of Gasser and Leutwyler four different powers of $N$ occur:
$N^k$ with $k=-2,-1,0,1$. Let us disregard for the moment the negative
powers -- we shall discuss them later. If we consider the case of $N$
degenerate quarks, the interpretation of the terms
linear in $N$ and constant is rather straightforward: they come from
the pion loops that contain one quark loop or none at all,
respectively. This conclusion is based on the simple observation that
since the strong interaction is flavour--blind, each quark loop
produces a factor $N$. Then, if quenched CHPT is working properly, 
the fermionic ghost loops have to cancel only the pion loops that give 
rise to a divergence proportional to $N$. Our main aim is to verify that 
this is exactly what happens.

For the sake of clarity we are going to ignore the presence of
the $U(1)_A$ anomaly of QCD, in the following. Taking it into
account would produce, as a main effect, a double pole in the
$\eta^{\prime}$ \cite{qCHPT}. How to cope with the diseases of this
part of the theory has been extensively analyzed by Bernard and
Golterman \cite{qCHPT,bg} looking at various different observables. 
In the perturbative treatment of the path integral we are considering,
these contributions simply add to the standard pion loop
contributions, and we can safely put them aside for the moment  \cite{long}.

The starting point of our calculation is the lowest order qCHPT 
Lagrangian, a graded symmetry generalization of the CHPT Lagrangian. 
This can be written as \cite{S}:
\begin{eqnarray}
{\cal L}^{\mbox{\tiny{qCHPT}}}_2&=&\frac{F^2}{4} 
\mbox{str} \left( D_\mu U_s D^\mu U_s^\dagger +
\chi^\dagger_s U_s + U_s^\dagger\chi_s \right) \; \; ,
\label{L2}
\end{eqnarray}
where str stands for the supertrace, the generalization of the trace
to graded matrices (see Ref. \cite{qCHPT} for details), 
$D^\mu U_s = \partial^\mu U_s - i r_s^\mu U_s+iU_s l_s^\mu$, $\chi_s=2
B_0(s_s+ip_s)$, and the field $U_s$ can be written as
$U_s=\exp(\sqrt{2}i/F\Phi)$, where 
$\Phi$ is the hermitian non traceless block--matrix 
\[ 
\Phi = \left( 
\begin{array}{cc}
\phi & \theta^\dagger  \\
\theta & \tilde\phi
\end{array}  \right)  
\] 
containing the physical pseudoscalar field $\phi$, the ghost field 
$\tilde\phi$, both of bosonic nature, and the hybrid fields
$\theta,\; \theta^\dagger$ of fermionic nature. The scalar field $s_s$
contains the quark mass matrix ${\cal M }$ through: 
\[
s_s=\left( 
\begin{array}{cc}
{\cal M } & 0  \\
0 & {\cal M}
\end{array}  \right)  + \delta{s_s} \; \; 
\] 
and that for our purposes is taken diagonal: ${\cal M} = m_q {\bf 1}$. 
All the Goldstone bosons will then have the same mass: $M^2=2 B_0m_q$. 
The Lagrangian in Eq. (\ref{L2}) contains the external fields $r_s^\mu,\;
l_s^\mu,\; s_s,\; p_s$, which are generalizations of the
standard external fields, in order to make the Lagrangian locally
invariant under the group $U(N|N)_R \times U(N|N)_L$. However, since we
are not interested in Green functions of the spurious fields, we directly
set to zero all the corresponding spurious external
fields. The generating functional will then become a function of the
usual external fields only: $r_\mu=v_\mu+a_\mu,\;l_\mu=v_\mu-a_\mu,\; s,\;
p$, which are all $N \times N$ matrices. 

To calculate quantum corrections we expand the leading order action 
given by the Lagrangian in Eq. (\ref{L2}) in the vicinity of the
classical solution which  
is determined by the external sources through the classical equations 
of motion. We define the classical solution as  $\bar{U_s}\equiv u_s^2$
and describe the fluctuations around it as $U_s=u_s \exp
(i\xi_s)u_s=u_s(1+i \xi_s-1/2 \xi_s^2 +\ldots )u_s$. To get the
generating functional to one loop we need to expand the action in the
$\xi_s$ field up to second order. Skipping all the details \cite{long} 
we end up with the following gaussian integral (in Minkowski space-time):
\begin{eqnarray}
e^{i Z^{\mbox{\tiny{qCHPT}}}_{\mbox{\tiny{one~loop}}}}&=& 
\int~d\mu\left[U_s\right]~\exp \left\{
i \int~dx {F^2 \over 4} \left[
\xi_a D^{ab}\xi_b +2 \zeta^\dagger_a \bar{D}^{ab}\zeta_b
+ \tilde\xi_a(\Box + M^2) \tilde\xi_a \right] \right\}\nonumber\\ 
&=&{\cal N} \frac{\det\bar{D}}{\left(\det D\right)^{1/2}}  \; \; ,
\label{DET}
\end{eqnarray}
where we used the decomposition of the fields $\xi ,\tilde\xi ,\zeta ,
\zeta^\dagger$ (quantum fluctuations corresponding to the classical
fields $\phi, \; \tilde\phi,\; \theta, \; \theta^\dagger$,
respectively) in terms of the $N^2$ generators of each $U(N)$ flavour
subgroup as $\xi = \xi^a{\hat\lambda_a}$,
with $\hat\lambda_a=\lambda_a/\sqrt{2}$ for $a=1,\ldots,\; N^2-1$ 
(the $\lambda_a$'s are the usual Gell--Mann matrices of $SU(N)$),
and $\hat\lambda_0={\bf 1}/\sqrt{N}$. We remark that the scalar
ghosts $\tilde\xi$ decouple from the physical pions and the integral over
them produces only an irrelevant constant. We also stress that both
differential operators $D^{ab}$ and $\bar{D}^{ab}$ are now functions only 
of the standard external fields and the $\phi$ field at the classical 
solution, since we have put to zero all the spurious external fields. 
The differential operator $D^{ab}$ is defined as follows 
\begin{eqnarray}
D^{ab}\xi_b &=& -d_\mu d^\mu\xi^a+\hat{\sigma}^{ab}\xi_b \nonumber\\
d_\mu\xi^a &=& \partial_\mu\xi^a+\hat{\Gamma}_\mu^{ab}\xi_b \; \; ,
\label{D}
\end{eqnarray}
where
\begin{equation}
\hat{\Gamma}_\mu^{ab}=-\langle \Gamma_\mu
[\hat\lambda^a,\hat\lambda^b ] \rangle \; \;, \; \; \; \;
\hat{\sigma}^{ab}= {1\over 4} \langle [u_\mu ,\hat\lambda^a ] 
[u^\mu ,\hat\lambda^b ] \rangle -{1\over 4} \langle
\{\hat\lambda^a,\hat\lambda^b\}\chi_+ \rangle \; \; .
\end{equation}
Let us recall some standard definitions in CHPT:
$\Gamma_{\mu }= 1/2 ( [u^\dagger ,\partial_\mu u ]-i u^\dagger
r_\mu u -i u l_\mu u^\dagger )$, $u_\mu= i u^\dagger D_\mu \bar{U}
u^\dagger$, and $\chi_+= u^\dagger \chi u^\dagger+u \chi^\dagger u$.
The differential operator $\bar{D}^{ab}$ acting on the ghost field $\zeta$
is defined like in Eq. (\ref{D}), but with barred quantities, given by
\begin{equation}
\bar{\Gamma}_\mu^{ab} =  -\langle\Gamma_\mu \hat\lambda^a \hat\lambda^b
\rangle \; \;, \; \; \; \;
\bar{\sigma}^{ab}=  -{1\over 4} \langle( u_\mu u^\mu +\chi_+ +4B_0{\cal
M}) \hat\lambda^a
\hat\lambda^b \rangle  \; \; ,
\end{equation}
where ${\cal M}$ is again the quark mass matrix, which is also contained 
in the external scalar field $s={\cal M}+\delta{s}$. 

The divergent part of Eq. (\ref{DET}) can be derived in closed form by
regularizing the determinants in $d$ dimensions and using standard heat 
kernel techniques. The result reads: 
\begin{eqnarray}
{i \over 2} \ln \det D &=& \frac{-1}{(4\pi)^2(d\!-\!4)} \int \! dx \left\{
{N \over 6} 
\langle \Gamma_{\mu \nu} \Gamma^{\mu \nu} \rangle + {1 \over 2}\left[
{1 \over 4} \langle u_\mu u_\nu \rangle \langle u^\mu u^\nu \rangle +
{1 \over 8} \langle u_\mu u^\mu \rangle^2 \right. \right. \nonumber \\
 &+&  {N \over 8} \langle (u_\mu u^\mu)^2 \rangle  + {N \over 4}
\langle  u_\mu u^\mu \chi_+ \rangle +  
{1 \over 4} \langle  u_\mu u^\mu \rangle \langle \chi_+ \rangle 
\nonumber\\
 &+& 
\left({N \over 8}-{1 \over {2N}} \right) \langle \chi_+^2 \rangle + 
\left({1 \over 8}+{1 \over {4N^2}} \right) \langle \chi_+ \rangle^2
 \nonumber\\ 
 &-& \left.\left. {1\over 2}\langle  u_\mu \rangle\langle u^\mu \left(
u_\nu u^\nu + 
\chi_+ \right) \rangle + {1 \over {2N}} \langle \chi_+^2 \rangle -
{1 \over {4N^2}} \langle \chi_+ \rangle^2 \right] \right\} +\ldots
\label{lndet1}
\end{eqnarray}
\begin{eqnarray}
 i\ln \det \bar{D} \! &=& \! \frac{-1}{(4\pi)^2(d\!-\!4)} \! \int \! dx 
\left[ {N \over 6}
\langle \Gamma_{\mu \nu} \Gamma^{\mu \nu} \rangle +{N \over 16}
\langle (u_\mu u^\mu + \chi_+ + 4 B_0 {\cal M} )^2 \rangle \right]
\!+\!\ldots 
\label{lndetg}
\end{eqnarray}
and their difference gives
\begin{eqnarray}
Z^{\mbox{\tiny{qCHPT}}}_{\mbox{\tiny{one~loop}}} &=&
\frac{-1}{(4\pi)^2(d\!-\!4)} \int \! dx \left\{ 
{1 \over 8} \langle u_\mu u_\nu \rangle \langle u^\mu u^\nu \rangle +
{1 \over 16} \langle u_\mu u^\mu \rangle^2 
- {1\over 4}\langle  u_\mu \rangle \langle u^\mu u_\nu u^\nu \rangle
\right. \nonumber\\
& & \; \; \; \; \; \; \; \; \; \; \; \; \; \; \; \; \; \; \; \; \; \;
\left. -{1\over 4}\langle u_\mu \rangle\langle u^\mu \chi_+ \rangle + 
{1 \over 8} \langle  u_\mu u^\mu \rangle \langle \chi_+ \rangle
 + {1 \over 16} \langle \chi_+ \rangle^2 
\right. \nonumber\\
& & \; \; \; \; \; \; \; \; \; \; \; \; \; \; \; \; \; \; \; \; \; \;
\left.  - {N \over 4} M^2 \langle u_\mu u^\mu \rangle 
- {N \over 4} M^2 \langle \chi_+ \rangle \right\} +
\ldots \; \; .
\label{Z}
\end{eqnarray}
The ellipses stand everywhere for the finite contributions to the one loop
generating functional. Eq. (\ref{Z}) shows explicitly the flavour 
dependence of the qCHPT functional to one loop and the comparison with
Eq. (\ref{lndet1}) provides the result we were after. 
We stress a few important points:
\begin{enumerate}
\item
Last line of Eq. (\ref{lndet1}) contains extra contributions with 
respect to the $SU(N) \times SU(N)$ determinant of Gasser and Leutwyler 
\cite{gl85}, due to the presence of the singlet field. Two of those terms 
are proportional to $\langle u_\mu \rangle \equiv\nabla_\mu \phi_0$, and
only contribute to processes with external singlet fields. 
The last two terms arise from loops with the singlet field running inside. 
These terms exactly remove the negative power dependence upon $N$ in the 
rest of the expression. This shows that the presence of such terms in 
the $SU(N)$ theory can be understood as an effect of the $U(1)_A$ anomaly 
that decouples the singlet field. 
\item
The $\zeta$ field loop in Eq. (\ref{lndetg}) produces only contributions 
linear in $N$ as expected, and cancels completely the terms linear in $N$ 
of the full generating functional. We also note that the $N$ dependence 
in Eq. (\ref{lndet1}) is not fully explicit. In fact, since the expansion 
of $\chi_+$ in powers of $\phi$ starts with a constant term proportional 
to the quark mass matrix, its trace is proportional to $N$ in the 
degenerate case we are considering here. The last two terms in 
Eq. (\ref{Z}) cancel this dependence in the final expression.
This result shows that quenched CHPT does what it should do, i.e. it
contains only the pion loops that do not contain quark loops.
\item
Each pole at $d=4$ produces a chiral logarithm after renormalization. 
A different way to present our result is to say that
we have calculated the chiral logs of any Green 
function in qCHPT to one loop. We obtained that these are nonzero in
general, though certainly different from those of the ordinary CHPT
case. How different has to be investigated case by case, since it is
not possible to identify a general behaviour from our result.
\end{enumerate}
\noindent In particular, from Eq. (\ref{Z}) it is easy to verify the 
results already obtained by Bernard and Golterman \cite{qCHPT} about the
absence of chiral logs in the chiral condensates, pion masses and
decay constants. 
The situation is different when one considers Green functions with
more than two external legs. To see what happens in one concrete
example, we consider the $\pi \pi$ scattering amplitude -- also
because there are lattice calculations of the two $S$--wave scattering
lengths available, Ref. \cite{fuku}.
As we argued above, the presence of chiral logs even in the quenched
theory has to be interpreted as due to diagrams with pion loops that
do not contain quark loops. For the $\pi \pi$ scattering amplitude an
example is given in Fig. \protect\ref{fig1}.

The complete amplitude in quenched CHPT reads:
\begin{eqnarray}
A_{\pi \pi}^{\mbox{\tiny{qCHPT}}}(s,t,u) &=& \frac{s-M_{\pi}^2}{F_{\pi}^2}
+ \frac{1}{4F_{\pi}^4} \left\{- \frac{1}{2} \left(L+{1\over 16\pi^2}\right)
\left[3 s^2 +(t-u)^2 \right]
 \right. \nonumber  \\
&+&  s^2 \bar{J}(s)+\left(t-2M_\pi^2\right)^2
\bar{J}(t)+ \left(u-2M_\pi^2\right)^2 \bar{J}(u) \nonumber \\
&+& \left. c_1 M_\pi^4+c_2 s M_\pi^2 +c_3^r(\mu) s^2 +
c_4^r(\mu) (t-u)^2 \right\} \nonumber \\
 &+& S_{(m_0,\alpha)} (s,t,u) +O(p^6) \; \; ,
\label{PPR}
\end{eqnarray}
where $\bar{J}(q^2)=J(q^2)- J(0)$, with
\begin{equation}
J(q^2) = \frac{1}{i}\int \frac{d^dl}{(2\pi)^d} \frac{1}{\left(M^2-l^2\right)
\left(M^2-(l-q)^2 \right)} \; \; ,
\end{equation}
and $L=(16 \pi^2)^{-1}$
$\log \left(M_\pi^2/\mu^2 \right)$.
The polynomial part with coefficients $c_1,\ldots c_4$ comes from the
counterterm Lagrangian at order $p^4$.
$S_{(m_0,\alpha)} (s,t,u)$ is the renormalized contribution to the 
amplitude of the singlet loops with one ($m_0$ and $\alpha$) vertex 
insertion on the singlet propagator. The effects and diseases of 
$S_{(m_0,\alpha)} (s,t,u)$ have been discussed in \cite{bg,pipiq}. 
$F_\pi$ and $M_\pi$ are the renormalized quenched values of $F$ and $M$ at
order $p^4$. While $F$ remains unchanged to one loop, $M$ gets a
one loop correction from  ($m_0$ and $\alpha$) vertex insertion which we
included in $S_{(m_0,\alpha)} (s,t,u)$. Finite contributions to
both $F_\pi$ and $M_\pi$ from the order $p^4$ Lagrangian are  
meant to be included in the renormalized polynomial part.
Let us stress that the divergent part of the amplitude before
renormalization, and the corresponding chiral logs in the renormalized
amplitude Eq. (\ref{PPR}) can also be obtained from the
generating functional, Eq. (\ref{Z}), by expanding it in powers of the
external fields for $N=2$, and taking the coefficient of the relevant
term. This offers a welcome check on the full calculation.

We focus now on the $S$--wave scattering lengths. In full CHPT it
is well known that the chiral logs dominate the one loop correction
 at $\mu = 1$ GeV \cite{gl83} . In the quenched case, the
coefficients of the chiral logs are:
\begin{eqnarray}
a_0^{0\;\mbox{\tiny{qCHPT}}}&=&\frac{7 M_\pi^2}{32 \pi F_\pi^2}
\left\{1+\frac{M_\pi^2}{F_\pi^2} \left(-\frac{22}{7}L + \ldots \right)
+ O(M_\pi^4) \right\} \; \; , \\ 
a_0^{2\;\mbox{\tiny{qCHPT}}}&=& -\frac{ M_\pi^2}{16 \pi F_\pi^2}
\left\{1+\frac{M_\pi^2}{F_\pi^2} \left(2 L + \ldots \right)
+ O(M_\pi^4) \right\} \; \; ,
\end{eqnarray}
while in full CHPT the corresponding coefficients are: $-9/2$ for
$I=0$ and $3/2$ for $I=2$  \cite{gl83}. In both cases 
the change in the correction due to quenching is of the order of 30\%; 
however the one loop correction itself is roughly of this size in full
CHPT  (see Ref. \cite{gl83}), so that the overall relative change can
be estimated to be around 10\%. 
Of course this is not the whole story: we have not included the
remaining analytic contributions from the regular part of the
amplitude, the constants $c_i$'s in Eq. (\ref{PPR}), and the
contributions from the ill part of the amplitude,
$S_{(m_0,\alpha)}(s,t,u)$. 
As for the first point, if one assumes that quenching will not change
the order of magnitude of the finite part of the $O(p^4)$ constants at
$\mu=1$ GeV (one could argue that some kind of Vector Meson Dominance
should be valid also in the quenched case), one can conclude that
those contributions should still produce a small correction.

The ill part of the amplitude is in principle more dangerous and requires 
a more careful treatment, as was done by Bernard and Golterman
\cite{bg}. They have shown in fact that this part dramatically changes
the very method to compute the scattering lengths on the lattice: the 
L\"uscher's formula \cite{luescher} that relates finite volume effects
to the $S$--wave scattering lengths is modified by quenching. 
However, the numerical analysis of this effect, adapted to the staggered 
fermions calculation by Fukugita et al. \cite{fuku} (the only
calculation done with a pion mass small enough to justify the
application of CHPT) indicates that this effect should be small.
Putting the various pieces together, we can conclude that a complete
analysis in qCHPT to one loop of the
$\pi \pi$ scattering amplitude, seems to provide an explanation for
the rather good agreement between the lattice calculation of $S$--wave
scattering lengths done in Ref. \cite{fuku} and standard CHPT,
as was pointed out by one of us in Ref. \cite{g}.

Let us summarize our results: we have calculated the divergent part of
the generating functional of quenched CHPT to one loop,
Eq. (\ref{Z}), and found that, for degenerate quark masses, {\em it does
not contain any explicit flavour number dependence}. 
Arguing that for a non--anomalous theory this dependence may arise only 
through the presence of quark loops (become manifest here through pion 
loops), we conclude that quenched CHPT is working properly. 
In our opinion this result puts on a very solid basis
quenched CHPT, the effective theory proposed by Bernard and Golterman
\cite{qCHPT} to describe the low energy physics of quenched QCD.
Our calculation provides also {\em the chiral logs of any Green function, 
and shows that these are nonzero in general}. As an example we have
discussed the case of the $\pi \pi$ scattering amplitude and found
that the coefficient of the chiral log of the two $S$--wave scattering
lengths has been modified by 30\%, 
a rather modest effect.

\section*{Acknowledgements}
It is a pleasure to thank G.M. de Divitiis, J. Gasser, M. Masetti and 
R. Petronzio for interesting discussions. We also thank J. Gasser and 
R. Petronzio for reading the manuscript.



\begin{thebibliography}{99}
\bibitem{S}
We always use the subscript $s$ to denote the graded symmetry version
of a standard CHPT quantity.
\end{thebibliography}
\end{document}